\documentclass{emulateapj}
\usepackage{apjfonts}
\usepackage{rotating}

\def\gax{\mathrel{\raise.3ex\hbox{$>$}\mkern-14mu\lower0.6ex\hbox{$\sim$}}}
\def\lax{\mathrel{\raise.3ex\hbox{$<$}\mkern-14mu\lower0.6ex\hbox{$\sim$}}}
\def\gtorder{\mathrel{\raise.3ex\hbox{$>$}\mkern-14mu
             \lower0.6ex\hbox{$\sim$}}}
\def\ltorder{\mathrel{\raise.3ex\hbox{$<$}\mkern-14mu
             \lower0.6ex\hbox{$\sim$}}}

\newcommand{\flux}{\hbox{erg~cm$^{-2}$~s$^{-1}$}}
\newcommand{\cmsq}{\hbox{cm$^{-2}$}}
\newcommand{\nh}{\hbox{${N}_{\rm H}$}}
\newcommand{\sarc}{$^{\prime\prime}\!\!.$}

\begin{document}

\title{SN~2002bu -- Another SN~2008S-like Transient}

\author{D.~M. Szczygie{\l}$^{1}$, C.~S. Kochanek$^{1,2}$, X. Dai$^{3}$}
 
\altaffiltext{1}{Department of Astronomy, The Ohio State University, 140 W. 18th Ave., Columbus OH 43210}
\altaffiltext{2}{Center for Cosmology and AstroParticle Physics, The Ohio State University, 191 W. Woodruff Ave., Columbus OH 43210}
\altaffiltext{3}{Department of Physics and Astronomy, University of Oklahoma, 440 W. Brooks Street, Norman, OK 73019}

\begin{abstract}

\noindent We observed SN~2002bu in the near-IR with the Hubble Space Telescope,
the mid-IR with the Spitzer Space Telescope and in X-rays with Swift 10 years
after the explosion.  If the faint $L_H \sim 10^2 L_\odot$ HST near-IR source at the transient position
is the near-IR counterpart of SN~2002bu, then the source has dramatically faded
between 2004 and 2012, from $L \simeq 10^{6.0}L_\odot$ to
$L \simeq 10^{4.5}L_\odot$.  
It is still heavily obscured, $\tau_V \simeq 5$ in
graphitic dust models, with almost all the energy radiated in the mid-IR.  
 The radius
of the dust emission is increasing as $R \propto t^{0.7 \pm 0.4}$ and
the optical depth is dropping as $\tau_V \propto t^{-1.3\pm0.4}$. The
evolution expected for an expanding shell of material, $\tau_V \propto t^{-2}$, 
is ruled out at approximately $2\sigma$ while the $\tau_V \propto t^{-0.8}$ to $t^{-1}$ optical 
depth scaling for a shock passing through a pre-existing wind is consistent
with the data.  If the near-IR source is a chance superposition, the present day 
source can be moderately more luminous, significantly more obscured and evolving
more slowly.  While we failed to detect X-ray emission, the X-ray flux limits
are consistent with the present day emissions being powered by an expanding
shock wave. SN~2002bu is clearly a member of the SN~2008S class of transients,
but continued monitoring of the evolution of the spectral
energy distribution is needed to conclusively determine the nature of the
transient.
\end{abstract}


\section{Introduction}
\label{sec:introduction}

Supernova (SN) 2002bu was discovered on 2002 March 28 by \cite{Puckett2002}
in the galaxy NGC~4242, with a fairly low peak magnitude $M_V \simeq -15$
\citep{Hornoch2002}. The last pre-discovery observations on 2001 February 21
and March 14 placed a unfiltered magnitude limit at the location of the  SN of
20.5~mag. A low resolution spectrum taken by \cite{Ayani2002} on April 1 showed
a flat continuum and strong, narrow Balmer emission lines (FWHM $\simeq$
1100 km/s) that led to a Type~IIn classification of the SN. The early ($\sim$80
days post peak) light curve presented by \cite{Foley2007} and \cite{Smith2011}
showed a plateau and a slower rise and decline as compared to other Type~II SNe,
with the color becoming redder with time.

\cite{Thompson2009} proposed that SN~2002bu may be a member of a new class
of stellar transients, with the prototypes being SN~2008S \citep{Arbour2008}
and the 2008 optical transient (OT) in NGC~300 \citep{Monard2008}. The class is
characterized by a transient--progenitor pair, where the transient is $2-3$ mag
fainter than a regular core collapse SN (ccSN), with narrow emission lines and
evidence of internal extinction in the spectrum, while the progenitor is a dust
enshrouded, low luminosity ($\sim$5$\times 10^4 L_{\odot}$) star with little
variability in the years before the outburst (\citealt{Prieto2008a}, \citealt{Prieto2008b}). 
In mid--IR color--magnitude
diagrams (CMD), the progenitors occupy the extreme end of the asymptotic giant
branch (AGB) sequence and the obscuring dust is graphitic rather than
silicate (\citealt{Prieto2009}, \citealt{Wesson2010}). 
\cite{Thompson2009} further showed that these transients are relatively common
among ccSNe ($\sim$20\%), but their progenitors are extremely rare
($\ltorder 10^{-4}$ of massive evolved stars), which implies that many massive
stars go through this dust obscured phase shortly ($\leq10^{4}$ yrs) before the
explosion.  The rarity of the progenitor stars was further confirmed in the
survey of additional galaxies by \cite{Khan2010}.

Since the main characteristic of the new class is a dust enshrouded progenitor,
it is impossible to unambiguously classify SN~2002bu as a member, as there are
no pre--explosion IR observations of the region. However, \cite{Thompson2009}
found a bright, red mid-IR source at the location of SN~2002bu in Spitzer Space
Telescope (SST) data taken 2 years after the explosion, which indicates dust
formation. This, together with the transient characteristics (low luminosity,
dust and narrow emission lines visible in the spectrum), make SN~2002bu
a likely member of the new class.
In addition, \cite{Smith2011} analyzed 5 spectra taken between 11 and 81 days
after discovery, and they show that the spectrum of SN~2002bu becomes redder
with time and evolves from a spectrum resembling an LBV in outburst at early
times, to one more similar to the spectra of SN~2008S and the NGC~300~OT 81 days
later. The evolution of the H$\alpha$ line from a Lorentzian profile to an
asymmetric, blueshifted Gaussian profile, is suggestive of newly formed dust.
In \cite{Kochanek2012} we argued that the early light curve also indicates 
dust formation, albeit with some ambiguities. 

In \cite{Kochanek2012} we analyzed archival Hubble Space Telescope (HST) and SST data from approximately
two years after the transient peak.  As previously reported by
\cite{Thompson2009}, SN~2002bu was a luminous mid-IR source in 2004, but
\cite{Kochanek2012} also found that it was invisible in archival HST data from
2005 in the BVRI bands to limits of $\sim 25$~mag.  The spectral energy
distribution (SED) in 2004 was well fit by surrounding a $T_{*}=20000$~K,
$L_{*}\simeq10^{5.9} L_{\odot}$ source with an optically thick
($\tau_{V}\simeq30$) shell of dust located at a radius of approximately
$R\simeq 10^{15.8}$~cm. 

\begin{figure}
\plotone{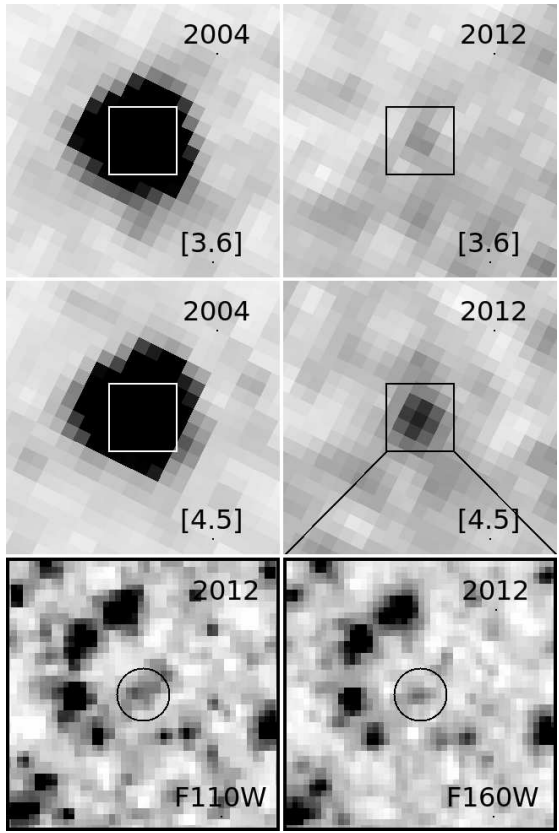}
\caption{The top two panels show the [3.6] $\mu$m SST observations
  of SN~2002bu from April 2004 (left) and January 2012 (right),
  the middle two panels show the corresponding [4.5] $\mu$m SST
  data, and the bottom two panels show the  F110W ($J$) and
  F160W ($H$) HST   observations from February 2012 of the region
  marked with a square on the SST images.
  The four top panels are 12\farcs0 $\times$ 12\farcs0. The
  two bottom panels are 3\farcs0 $\times$ 3\farcs0 (about 80 $\times$
  80 parsecs) and the radius of the circle is $\sim$0\farcs3,
  which is 3 times the uncertainty in the astrometry.
}
\label{fig:ssthst}
\end{figure} 

Broadly speaking, there are three possible explanations of the observations, as
we discussed in our mid-IR survey of the ``supernova impostors''(\citealt{Kochanek2012}).
The first, ``traditional'', view of these events is that a shell of material is
ejected during the optical transient and forms dust once the ejected material
becomes cool enough.  As the shell expands, its optical depth drops as
$\tau \propto 1/t^2$ and the characteristic temperature drops as
$T \propto L(t)^{1/4} t^{-1/2}$ because the shell expands with the ejecta
velocity $R \simeq v_s t$.  The dust radius at the time of the previous SST
observations of SN~2002bu was consistent with a shell expanding at the velocity
of $v_s=893$~km/s adopted by \cite{Smith2011}.  
The second possibility is that the optical transient is a signal that the star
is entering a high mass loss phase with a dense wind that forms dust and
obscures the source.  While the wind is steady, the optical depth is roughly
constant with a dust temperature close to the dust destruction temperature
$T_d \simeq 1500$~K unless the wind becomes optically thick in the mid-IR.
When the high mass loss phase ends, the evolution quickly resembles the first
scenario (see \citealt{Kochanek2012}).
The third scenario is the one introduced by \cite{Kochanek2011} to explain
SN~2008S and the NGC~300~OT.   Here the progenitors are already shrouded by
a very dense wind when an explosive transient occurs that destroys most of the
dust to leave the transient little obscured at peak.  The wind is so dense, however,
that the dust reforms and re-obscures the transient.  The present day
luminosity is a combination of a surviving star (if any, nothing in the
data requires one) and the luminosity generated
by the shock propagating through the wind.  At later times, the optical depth
outside the shock is dropping as $\tau \propto 1/t$ and, once the optical depth
is low enough, the X-rays produced in the shock should be observable.
 
These scenarios make different predictions for the time evolution of the
transient, so monitoring the evolution of SN~2002bu as a function of wavelength
should reveal the nature of the event.  Here we report near-IR HST, mid-IR SST
and Swift X-ray observations of SN~2002bu taken roughly 10 years after the
transient peak and 8 years after the the last HST and SST observations.
Section~2 presents the new observational data.  We discuss the results in Section~3,
and consider their broader implications in Section~4.  

\section{Observations and Data Analysis}

\begin{figure}
\plotone{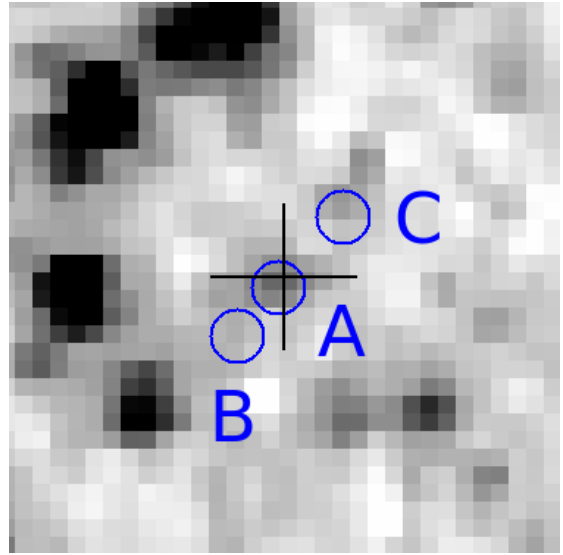}
\caption{A 2\farcs0 $\times$ 2\farcs0 region of the F160W HST
  images with a cross marking the estimated position of SN~2002bu.  
  The arms of the cross span 2$\sigma$ of the astrometric uncertainties.
  The circles mark the three sources detected by DOLPHOT within 
  the 3$\sigma$ error in the estimated SN location with A being  
  the closest and C the most distant.}
\label{fig:closeup}
\end{figure}

We observed SN~2002bu with SST in June 2011 and January 2012, in both the [3.6]
and [4.5] $\mu$m bands (program ID 80015) with exposure times of 240~sec
(8 dithered 30 sec exposures) for both bands.  We also observed it with HST in
February 2012 using the WFC3/IR camera and the F110W ($J$) and F160W ($H$)
filters (proposal ID 12450) with exposure times of 2$\times$700~s for each band.
Figure~\ref{fig:ssthst} shows the changes in the [3.6] and [4.5] images between
2004 and 2012 as well as a zoomed in view of the region in the $J$ and $H$-band
HST images.  The source has become significantly fainter in the mid-IR, more so
at [3.6] than at [4.5], but any near-IR counterpart must still be very faint.
Figure~\ref{fig:closeup} shows a close-up view of the region in the F160W image.
We identified 12 reasonably isolated stars in the HST images that could be
matched to the SST images using the IRAF {\em immatch} package.  Using the 2004
$3.6\mu$m image as the reference frame, we obtain the estimated SN position in the
HST data shown by the circle in Figure~\ref{fig:ssthst} and a cross in
Figure~\ref{fig:closeup}.  The uncertainty in the position is approximately
1.5~HST pixels or $0\farcs09$.

The Spitzer fluxes were measured with aperture photometry ({\em apphot} package
in IRAF), following the procedure described in \cite{Kochanek2012}.  We measure
the flux of the source using a range of source and sky apertures (with
appropriate aperture corrections) and then combine the individual results taking
into account both the formal statistical and the systematic uncertainties
implied by the scatter in the fluxes found for different aperture combinations.
We analyzed the HST images using the DOLPHOT \citep{Dolphin2000} photometry 
package with its standard WFC3 parameter files.  DOLPHOT identifies three
sources within a radius three times larger than our estimated astrometric
uncertainties.  We label these sources A, B and C in order of their 
distance from the estimated position (see Figure~\ref{fig:closeup}).  
The photometric results are presented in Table~\ref{tab:mags} together with
the previous HST and SST flux measurements from \cite{Kochanek2012}.

We adopt a distance to NGC~4242 of 5.8~Mpc based on \cite{Tully2009} and a foreground 
Galactic extinction of $E(B-V)=0.01$ (\citealt{Schlegel1998}).  We note, however, that
the distance to NGC~4242 is uncertain.  There is a second Tully-Fisher distance estimate 
of 10.4~Mpc by \cite{Springob2007} (erratum in \citealt{Springob2009}), and if
we associate NGC~4242 with the group containing NGC~4258 it lies at the intermediate
distance of $7.2$~Mpc (\citealt{Hernstein1999}).  Our qualitative conclusions are
unaffected by these uncertainties.  Quantitatively, luminosities increase proportional
to $d^2$, dust radii increase as $d$, and the velocities implied by the dust radii
increase as $d$.  The actual photometric models are essentially distance independent
other than these scalings.  We comment on the effects of distance changes as necessary.

\begin{figure*}[t]
\plotone{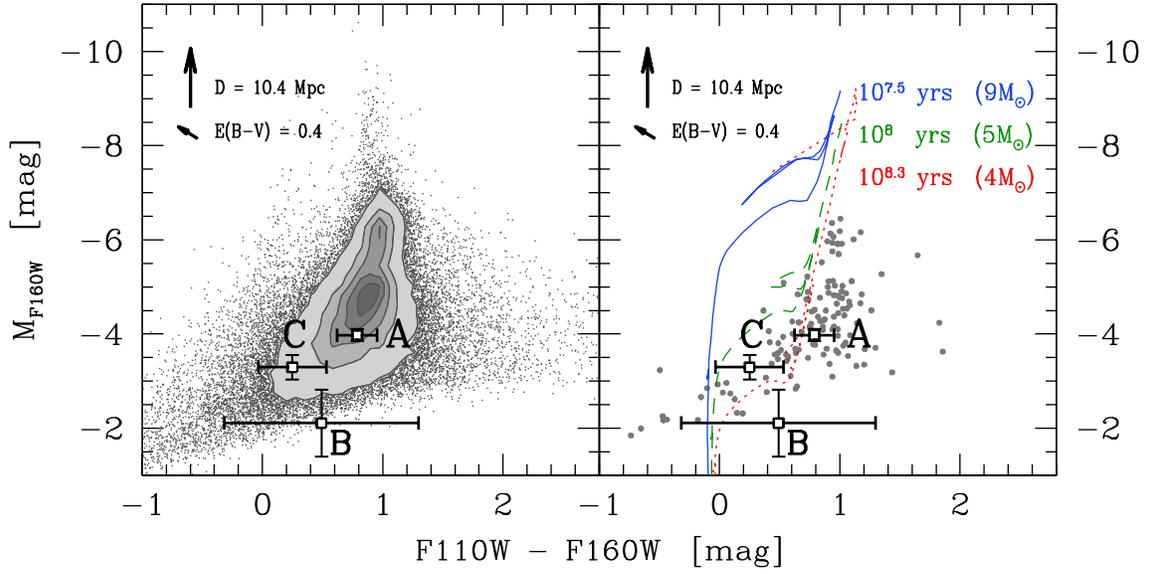}
\caption{Near-IR CMDs from the HST F110W and F160W observations,
  corrected for Galactic extinction of 0.012 and assuming a distance
  of $5.8$~Mpc \citep{Tully2009}.
  The left panel shows a large region (2\farcm3$\times$2\farcm0) while
  the right panel shows a $\sim$3\farcs5 (100 pc) radius around
  SN~2002bu. The locations of the three objects in the SN~2002bu
  region (A, B, and C, see Fig.~\ref{fig:closeup}) are marked with
  labeled squares.
  The curves on the right panel show Padova (\citealt{Marigo2008})
  isochrones for $10^{7.5}$, $10^{8.0}$ and $10^{8.3}$~years,
  alternating solid and dashed lines, which have end points
  corresponding to ZAMS masses of $9$, $5$ and $4M_\odot$ respectively.
  The vertical arrows shows how much the points would shift if instead
  we adopted the distance of 10.4~Mpc. The diagonal arrows show
  how the points would shift after correcting for $E(B-V)=0.4$~mag of
  additional extinction while keeping the distance of 5.8~Mpc.
}
\label{fig:cmd}
\end{figure*}

Figure~\ref{fig:cmd} shows near-IR H/J$-$H CMDs constructed from the DOLPHOT catalogs
for a large (2\farcm3$\times$2\farcm0) and a small 3\farcs5 (100 pc) radius region around 
the SN.  We required the signal-to-noise ratio in both filters to be greater than 4 for
a detection (see \citealt{Dalcanton2011}), and a sharpness parameter
$sharpness^2 < 0.1$ to exclude non-stellar sources. The CMDs are corrected for
Galactic extinction ($E(B-V)=0.012$~mag, \citealt{Schlegel1998}) and assume a
distance of $5.8$~Mpc \citep{Tully2009}. We mark the locations of sources
A, B and C in both CMDs. The right panel in Figure~\ref{fig:cmd}
also shows the Padova
(\citealt{Marigo2008}) isochrones for $10^{7.5}$, $10^{8}$ and $10^{8.3}$~years,
that have ZAMS masses corresponding to their end points of $9$, $5$ and
$4M_\odot$, respectively.  The isochrones are moderately bluer than the
stars, which could be evidence for $E(B-V) \simeq 0.4$~mag (for 5.8~Mpc,
$0.2$~mag for $10.4$~Mpc) of local extinction.  Unfortunately, the optical
detection limits are not strong enough to constrain these possibilities.
For all these distances and extinctions, the local stellar population is
only consistent with the older isochrones, ages closer to $10^8$ years
and a maximum mass of $5 M_\odot$ than $10^{7.5}$~years and a maximum mass
of $9 M_\odot$.  Since adding these extinctions have little effect on our
near/mid-IR models, we add no additional extinction beyond Galactic to the
models in \S3.

We also observed SN2002bu with \emph{Swift} \citep{Gehrels2004} between April
26--30 2011, for a net exposure of 15~ks. We reprocessed the XRT data using the
\verb+xrtpipeline+ tool provided by the Swift team and reprojected the
observations into a single image. We chose the source region to be a circle
centered on the SN with a radius of 10 pixels (23\sarc6) and a nearby
background region without any sources. We did not detect the source either in
the full (0.2--10~keV) or the 0.2--0.5~keV, 0.5--2~keV, and 2--10~keV bands.
We obtained 1$\sigma$ limits of $<5.4$, $<1.3$, $< 3.8$, and
$<3.4 \times10^{-4}$~cnt~s$^{-1}$, respectively, for these bands, where we
corrected the small aperture used in the analysis based on the \emph{Swift} PSF
\citep{Moretti2005}.
Assuming thermal bremsstrahlung emission with $T_X=0.8$~keV and Galactic
absorption of $\nh_{Galactic} = 1.17\times10^{20}~\cmsq$ \citep{Dickey1990},
we obtained a flux limit of $< 0.9\times10^{-14}~\flux$ in the 0.5--2~keV band
that roughly corresponds to a luminosity limit of $L_X < 10^4 L_\odot$. More
generally, we model the emission as thermal brehmsstrahlung emission at a
temperature $T_X$ obscured by an additional column density $\nh$ associated
with the source.  We use the PIMMS model for the Swift/XRT to estimate
the absorbed soft, medium and hard-band count rates and constrain the
unabsorbed total luminosity to be consistent with the observed upper limits
for the three energy bands.

\section{Results and Discussion}

\begin{figure*}
\plotone{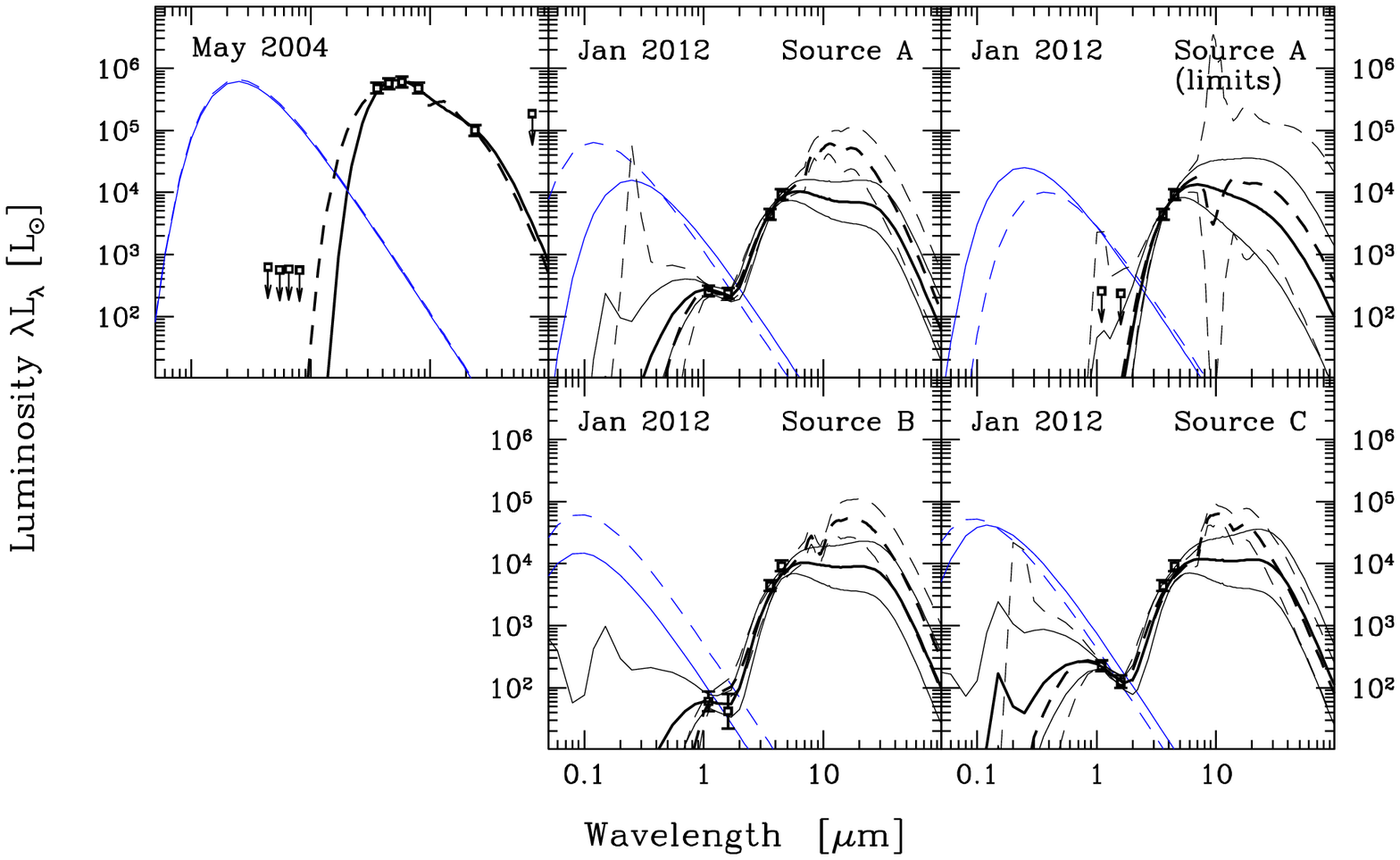}
\caption{The top left panel shows the SN~2002bu SED roughly 2
  years after the explosion, while the other 4 panels show the
  current SED for sources A, B, and C (see Fig.~\ref{fig:closeup}),
  in the case of source A also treating HST near-IR measurements
  as upper limits (top right panel).  Magnitudes from
  Table~\ref{tab:mags} are converted to fluxes, and then to
  luminosities as $L=4\pi D^2 \nu  F_\nu$ where $D=5.8$~Mpc.
  The SED models are black bodies with $A_V=0.012$~mag of total
  (Galactic) extinction. The thick black curves show the probability
  averaged SEDs for graphitic (solid) and silicate (dashed) dust,
  while the thin curves show the spread around the mean values.
  In the case of the 2004 SED, solid and dashed lines are the best
  fit graphitic and silicate models taken from \cite{Kochanek2012}
  and correspond to a $T_{*}=20000$~K, $L_{*}\simeq10^{5.9} L_{\odot}$
  star surrounded by an optically thick ($\tau_{V}\simeq30$) shell
  of dust expanding at $\sim$900~km/s.
}
\label{fig:sed}
\end{figure*}

We used DUSTY (\citealt{Ivezic1997}; \citealt{Ivezic1999}) to fit the SEDs
of the sources detected in HST near-IR data. Figure~\ref{fig:sed} shows the
probability averaged SEDs of SN~2002bu in May~2004 and January~2012
(for sources A, B and C, treating near-IR HST measurements of
source A either as detections or upper limits).
The best fit model to the 2004/2005 data is taken from \cite{Kochanek2012} and
corresponds to a $T_{*}=20000$~K, $L_{*}\simeq10^{5.9} L_{\odot}$ star
surrounded by an optically thick ($\tau_{V}\simeq30$) shell of dust expanding
at $\sim$900~km/s. 
Both graphitic and silicate models give similarly good fits
to the data.

We also embedded DUSTY in a Markov
Chain Monte Carlo engine to model the SEDs, varying the dust temperature,
$T_d$, optical depth, $\tau_V$, and stellar temperature, $T_*$, with either
a fixed 2:1 ratio between the inner and outer radii of the shell or allowing
the ratio to vary between 1.1 and 10.  We included a weak prior on the stellar
temperature, $\log_{10} T_* = 4.0 \pm 0.3$, (and restricted its range to
$3000~\hbox{K} < T_* < 30000~\hbox{K}$) and on the expansion velocity implied by the inner radius
of the shell, $\log_{10}(v_s/\hbox{km/s})=\log_{10}(893)\pm0.30=2.95\pm0.30$.  For the 2012 epoch we
ran models considering star A as either a detection or an upper limit.  The
results allowing for variations in the shell thickness were little different
from those with a fixed thickness and no particular thickness was preferred, so
we only report the results for the fixed thickness in Table~\ref{tab:results}.  

The results for the 2004 epoch are the same as in our earlier models from
\cite{Kochanek2012}.  The source must be quite luminous, 
$L_* \simeq 10^{5.92 \pm 0.02}L_\odot$, but with an indeterminate source 
temperature because of the heavy obscuration.  The optical depth in graphitic
models is $\log_{10} \tau_V \simeq 1.6 \pm 0.2$ (scattering plus absorption), and it
is moderately higher in the silicate models because of the higher scattering
opacities of silicate dusts (see the discussion in \cite{Kochanek2012}).
Significantly higher optical depths begin to have significant opacity even for
the shorter wavelength IRAC bands, inconsistent with the shape of the SED.
The dust temperatures at the inner edge of $T_d \simeq 1100 \pm 170$~K are
roughly in the range expected for newly forming dust. 
The dust radius is estimated to be $\log_{10} (R_{in}/\hbox{cm}) \simeq 15.77 \pm 0.12$,
implying a velocity of $\log_{10} (v_s/\hbox{km/s}) \simeq 2.95 \pm 0.12$ that is consistent
with the velocity prior taken from the line widths cited by \cite{Smith2011}.
Note, however, the uncertainties are much smaller than those of the prior, 
so the data are in fact determining this radius and velocity rather than the
prior.  Despite the reasonable coverage of the mid-IR SED, graphitic
and silicate dusts fit the data equally well.  As discussed earlier,
the primary effect of distance uncertainties is simply to rescale the
luminosities, velocities and distances.  In particular, adopting the
distance to NGC~4258 ($7.2$~Mpc) raises the velocity from $v_s \simeq 900$~km/s
to $v_s \simeq 1100$~km/s, while using the \cite{Springob2009} distance
of $10.4$~Mpc raises it to $v_s \simeq 1500$~km/s.  One reason we 
adopted the smaller distance is that the larger distances begin 
to require significantly larger mean expansion velocities than implied
by spectroscopic observations during the transient (see \citealt{Smith2011}).

The interpretation of the SED in 2012 depends critically on whether any source
in the HST images corresponds to SN~2002bu.  If source A (or B/C) is
a detection, then the present day source must be a relatively hot,
$T_* > 15000$~K, star.  For graphitic dust, the star is relatively low
luminosity $L_*=10^{4.5\pm0.2} L_\odot$, with moderate 
$\log_{10} \tau_V \simeq 0.70\pm0.25$ obscuration, at a relatively large radius
$\log_{10} (R_{in}/\hbox{cm}) \simeq 16.18\pm0.25$ that corresponds to a lower average expansion
velocity $\log_{10} (v_s/\hbox{km/s}) \simeq 2.72\pm 0.25$ but is uncertain enough to be
consistent with no change in velocity. The dust temperature at the inner edge
is much cooler, $T_d \simeq 500$~K. In the silicate models the source is more
luminous, $L_*=10^{4.95\pm0.18}L_\odot$ with greater optical depth,
$\log_{10} \tau_V \simeq 1.21\pm0.12$, located at a smaller radius
$\log_{10} (R_{in}/\hbox{cm}) \simeq 15.98\pm0.17$, that now requires a slower expansion rate
$\log_{10} (v_s/\hbox{km/s})\simeq 2.52 \pm 0.17$.  The dust  temperature at the inner edge is
moderately warmer, $T_d \simeq 600$~K. There is no basis for choosing between
the two dust types.
If the association with source A is simply a coincidence and we instead treat
the HST fluxes as upper limits, then solutions with cool stars are allowed,
but they require increasingly high optical depths for lower temperatures,
$\tau_V \simeq 100$ for the coolest ($T_*=2500$~K) models. This is essentially
the limit of the fully obscured progenitors of SN~2008S and the NGC~300-OT
(\citealt{Prieto2008a}, \citealt{Prieto2008b}, \citealt{Thompson2009}, \citealt{Kochanek2011}).

\begin{figure}
\plotone{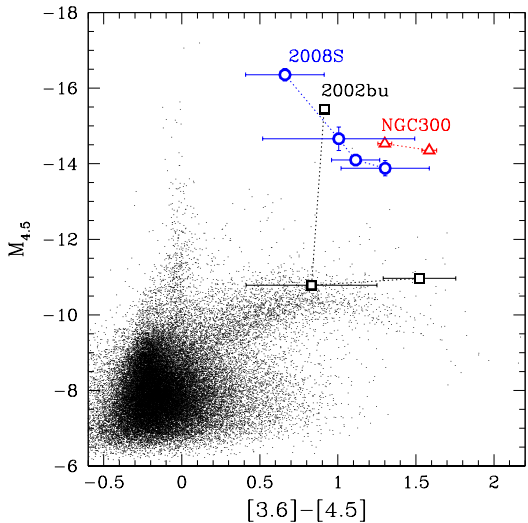}
\caption{The mid-IR CMD of NGC~2403 (\citealt{Khan2010}).
  The color evolution of SN~2002bu is marked with black squares
  connected by a dotted line, where the top point corresponds to
  the 2004 epoch, while the next to 2011 and 2012 epochs (see
  Table~\ref{tab:mags}).
  For comparison, we show the color evolution of SN~2008S from 2008
  to 2012 (blue circles)  and NGC~300-OT from 2008 to 2012 (red triangles)
  as summarized in \cite{Kochanek2011} and \cite{Szczygiel2012}.
}
\label{fig:midIR_cmd}
\end{figure}

The current position of the source in the mid-IR CMD (see Figure~\ref{fig:midIR_cmd}), 
close to the tip of the AGB sequence, is similar to that of the SN~2008S or NGC~300-OT 
progenitor stars (\citealt{Prieto2008a}, \citealt{Prieto2008b}).  This is simply
a coincidence driven by the present luminosity and dust radius, because in 
steady state a surviving star could not support the necessary optical depths
given the velocities needed to have new material at such large distances.
The population studies by \cite{Thompson2009} and \cite{Khan2010} imply
a lifetime in the obscured phase of order $t_o \simeq 10^4$~years.  The mass
loss rate is related to the optical depth by
\begin{equation}
         \tau_V = { \dot{M} \kappa_V \over 4 \pi v_w R_{in} },
       \label{eqn:opdepth}
\end{equation} 
for visual opacity $\kappa_V = 100 \kappa_2$~cm$^2$/g, mass loss rate
$\dot{M}$ and wind speed $v_w$, which implies a total mass loss of
\begin{eqnarray}
   \dot{M} t_o &\simeq  &4 \pi v_w R_{in} \tau_v t_o \kappa_V^{-1} \\
          &\simeq &{200\over \kappa_2} \left( { v_w \over 10^3~\hbox{km/s} }\right)
                  \left( { R_{in} \over 10^{16}~\hbox{cm} } \right)~
                  \left( { \tau_V \over 10 } \right)
                  \left( { t_o \over 10^{4}~\hbox{years} }\right)
               M_\odot \nonumber
\end{eqnarray}
while the star is in its obscured phase.  No star could sustain this for
$v_w \sim 10^3$~km/s, so the
current state cannot represent a long lived period and must be a transient
phase.  The progenitors, under the super-AGB star
hypothesis of \cite{Thompson2009}, can have the necessary lifetimes because
dust driven wind velocities of $v_w \simeq 10$-$20$~km/s are so low
(e.g. \citealt{Ivezic2010}).  However, such a slow wind commencing
after the transient would not even have started to form dust at this point
in time.

\begin{figure*}
\plotone{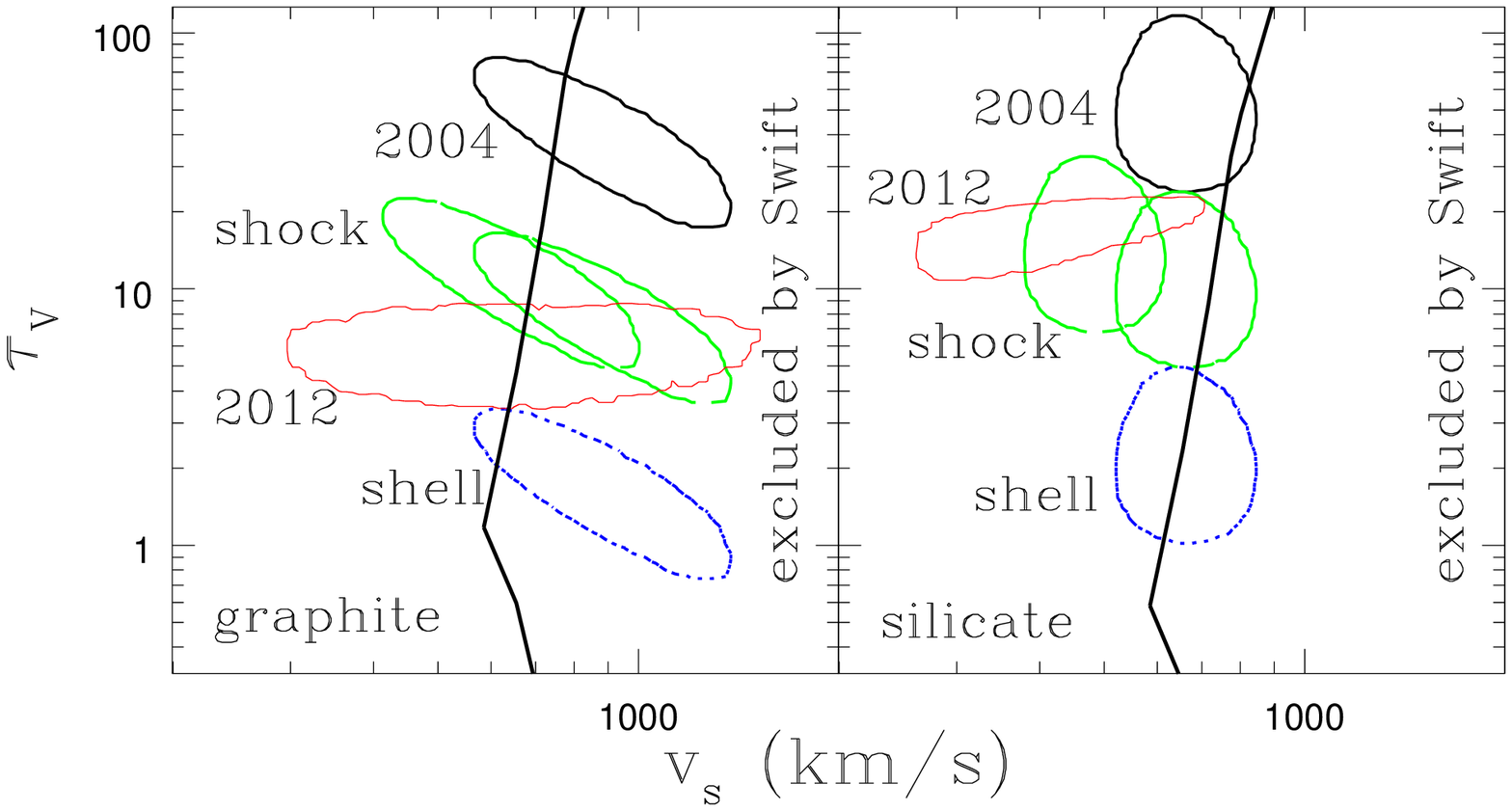}
\caption{The average expansion velocity and optical depth of the DUSTY models
  for SN~2002bu for graphitic (left) and silicate (right) dusts.  The 
  black (red) contours encompass the $1\sigma$ region for two parameters
  for the 2004 (2012) epoch.  The dashed green contours (labeled shock) show the 
  expected parameters in 2012 given the parameters in 2004 assuming the
  radius is expanding as $R\propto t^x$ with $x=0.8$ (left) or $1.0$ (right)
  and the optical depth is dropping as $\tau_V \propto 1/R$.  The dotted
  blue (labeled shell) shows the expected parameters in 2012 if the
  radius is expanding as $R \propto t$ and the optical depth is dropping
  as $\tau_V \propto 1/t^2$.  Shock models to the right of the heavy solid line 
  are excluded ($1\sigma$) by the upper limits on the X-ray flux in the Swift 
  observations.  If the HST sources are treated as upper limits, the 2012 
  contours expand upwards to higher optical depths.  The optical depth
  is related to the wind density by Eqn.~\ref{eqn:opdepth} where we 
  used $\kappa_2=1.7$ ($0.84$) for graphitic (silicate) dusts.
}
\label{fig:params}
\end{figure*}
            
The second possibility is the traditional view of the supernova impostors as
stars that briefly enter a high mass loss state during the transient to produce
a shell of ejected material that then forms dust.  As discussed in
\cite{Kochanek2012}, SN~2002bu probably started forming dust too soon after
the transient peak to have formed it in ejected material.  The more serious
problem is that between 2004 and 2012 an expanding, dense shell should have
a radius growing as $R_{in} \propto t$ and an optical depth dropping as
$\tau_V \propto t^{-2}$.  If we take the elapsed time from discovery to the
IRAC observations in 2004 (736 days) and 2012 (3566 days), the radius should
have increased by a factor of $4.9$ and the optical depth should have dropped
by a factor of $23.5$.  Fig.~\ref{fig:params} illustrates this by comparing the
mean expansion velocities and optical depths for the two epochs and both grain
types.  If we scale the values in 2004 to their expected values in 2012 simply
using the expected temporal scalings, we see that the observed properties in
2012 have a significantly higher optical depth and may require some
deceleration of the expansion rate.  

This leaves the scenario developed by \cite{Kochanek2011} for SN~2008S and
the NGC~300-OT, where the absorption is due to a dense pre-existing wind.
In this scenario, the progenitors were obscured by a dense, dusty wind where
the dust is destroyed by the shock breakout luminosity from the explosive
transient so as to leave the transient with little obscuration at peak.
The densities are so high, however, that the dust can then reform and
re-obscure the system, consistent with the early dust formation suggested
by the spectra and photometry.  In the SN~2008S scenario, the material already
exists at these distances and so dust can begin (re-)forming very rapidly.  If the
luminosity is then powered by a combination of any surviving star and an
expanding shock wave, the optical depth to the shock front is only dropping as
$\tau_V \propto 1/t$.
The resulting drop in the optical depth by only a factor of $4.9$ is far more
compatible with the observations, as shown in Fig.~\ref{fig:params}.  Arguably,
the shock should also be slowing as it expands through the dense wind. Self-similar solutions
(e.g. \cite{Chevalier1983}) give expansion rates of $R \propto t^s$ and
$v_s \propto t^{s-1}$ where $s=(n-3)/(n-2)$ for ejecta with an effective
density profile $\rho \propto R^{-n}$ expanding into a $\rho \propto R^{-2}$
wind. The typical approximations for the structure of the ejecta are $n=7$
($s=0.8$) and $n=12$ ($s=0.1$), leading to a modest slowing with time.
In Fig.~\ref{fig:params} we show the scaling of the conditions in 2004 to 2012
for this shock scenario ($v_s \propto t^{-0.2}$, $R \propto t^{0.8}$, 
$\tau \propto t^{-0.8}$).

If a significant part of the present day luminosity is driven by X-rays from
an expanding shock, inferences about the temperature of the illuminating source 
and the exact value of the optical depth become somewhat problematic.
In the shock scenario, the X-rays are absorbed by the
dense gas and remitted as a complex, non-thermal (mainly) emission line
spectrum, which is then absorbed and reradiated by the dust. While the shock
emission models of \cite{Allen2008} do not extend to the regimes considered
here, they generally have relatively low near-IR emission compared to
optical/UV emission and so would resemble a hot star model.  The
estimates of the optical depth, which are driven by the near-IR
detections, should also be viewed as being only logarithmically
correct in this scenario.   The overall luminosity, dust temperature
and dust radius, which are largely determined by the mid-IR emissions from
the dust, should still be accurate.

If we fit the expansion of the dust radius as a power law, $R \propto t^x$, we
find $x=0.67$ ($0.26 < x < 1.05$, $1\sigma$) for the graphitic models and 
$x=0.54$ ($0.29 < x < 0.80$, $1\sigma$) for the silicate models.  The
silicate models are consistent with linear expansion only at $2\sigma$.
Similarly, if we fit the evolution of the optical depth as a power law,
$\tau_V \propto t^y$, we find $y=-1.27$ ($-1.69 < y < -0.90$, $1\sigma$)
and $y=-0.79$ ($-1.29 < y < -0.47$, $1\sigma$) for the graphitic and
silicate models, respectively.  The $\tau_V \propto 1/t^2$ scaling of
expanding shells is ruled out at roughly $2\sigma$.  We can also fit 
the evolution of the optical depth as a power law in radius, $\tau_V \propto R^z$,
finding $z=-1.75$ ($-3.90 < z < -1.01$) and $z=-1.52$ ($-3.46 < z < -0.69$)
for the two models.  While this appears to be consistent with the $\tau_V \propto R^{-2}$
scaling for an expanding shell, there is a strong covariance between this exponent
and the expansion rate of the shell -- a solution with $\tau_V \propto R^{-2}$ 
must also roughly have $R \propto t^{1/2}$.  Essentially, for the same change
in optical depth there is considerable uncertainty in the radius because we
lack a complete dust SED in 2012 for constraining the dust temperature. In
the two-dimensional probability distribution of the $x$ and $z$ exponents,
the expanding shell solution with $x=1$ and $z=-2$ is still ruled out at 
roughly $2\sigma$.  The evolution is consistent with the $\tau_V \propto 1/R$
and $R \propto t^{0.8-1.0}$ evolution of the expanding shock model.

A shock moving through a wind at velocity $v_s$ has a characteristic X-ray
energy of 
\begin{equation}
    E_s = { 3 \mu \over 16 } m_p v_s^2 = 1.2 \left( { v_s \over 1000~\hbox{km/s}} \right)^2~\hbox{keV}
\end{equation}
where the mean molecular weight is $\mu=0.6$, and produces luminosity 
\begin{equation}
    L_s =  { \epsilon \over 2 } { \dot{M} \over v_w } v_s^3 
\end{equation}
where $\epsilon \leq 1$ is the efficiency with which the shock energy is
radiated as X-rays (e.g. \citealt{Chevalier1982}, \citealt{Chugai1992}, \citealt{Chugai1994},
\citealt{Chevalier1994}).  The radius of the shock 
$ R=v_s t =3.1 \times 10^{16} (v_s/1000~\hbox{km/s})$~cm 
is simply related to its velocity, where $t=3566$~days  is the
elapsed time from the transient to the X-ray observation.  This means that the
hydrogen column density outside the shock radius is 
\begin{eqnarray}
     N_H &= &{ \dot{M} \over 4 \pi v_w  R (1.4m_p) } \\
       &= &7.5 \times 10^{21} \left( { \dot{M} \over 10^{-4} M_\odot/\hbox{year} }\right)
        \left( { 10~\hbox{km/s} \over v_w} \right) 
        \left( { 1000~\hbox{km/s} \over v_s }\right)~\hbox{cm}^2 \nonumber
\end{eqnarray}
Combining these three equations with the PIMMS models for absorbed thermal
bremsstrahlung, we can determine the range of shock velocities and wind density
parameters ($\dot{M}/v_w$) that would violate the Swift limits, as shown in
Fig.~\ref{fig:params}.  If the shock velocity is as high as the $v_s = 893$~km/s
FWHM reported by \cite{Smith2011}, then there are allowed solutions for low wind
densities (where the X-ray luminosity is low but little absorbed), and high wind
densities (where the X-ray luminosity is high but heavily absorbed).  If the
shock velocity $v_s < 600$~km/s, then the X-ray emission limits are satisfied
for any wind density.  The parameters derived from the photometric fits broadly
satisfy these limits on the X-ray emission. Half the graphitic solutions
violate the limit in Fig.~\ref{fig:params}, but the limit as drawn is for 100\%
conversion of the shock energy into soft X-rays and a $1\sigma$ detection
threshold, which is a rather optimistic representation of the detection
threshold.  The basic picture of Fig.~\ref{fig:params} is little changed 
if we use the larger distances to NGC~4242, with all solutions and limits simply
shifting to higher velocities.  While this increases the shock luminosity,
the net effect is limited because the X-ray luminosity limits from the 
observations become correspondingly weaker.

\section{Summary}

Combining Spitzer, Hubble and Swift observations of SN~2002bu, we confirm the
arguments in \cite{Thompson2009} and \cite{Kochanek2012} that this source is
a member of the SN~2008S class of transients.  While we have no direct evidence
that the progenitor was self-obscured, dust appears to have begun forming
before any ejecta would be cool enough to do so and the optical depth of the
dust is dropping significantly more slowly than the $\tau_V \propto 1/t^2$
scaling of an expanding shell of material.  The evolution is broadly consistent
with the expanding shock model of \cite{Kochanek2011} for SN~2008S and the
NGC~300-OT.  It is not consistent with an expanding shell of ejected material
or a reformed, slow wind around a surviving dusty star. Unfortunately, our
Swift X-ray observations only constrain the allowed parameter space since we
failed to detect X-ray emission from the shock.

\cite{Gogarten2009} argued from an analysis of the stellar populations near
the NGC~300-OT that the progenitor could be as massive as $\sim 20M_\odot$,
while \cite{Prieto2008a} argued for $\sim 10M_\odot$ because the progenitor
appeared to be an extreme AGB star based on its location in mid-IR CMDs.
The ambiguity is possible because the most luminous AGB stars have 
luminosities comparable to evolved $ \sim 20 M_\odot$ stars.  Formally,
however, these population analyses only provide upper mass limits, so it
is entirely possible to have a transient (even a supernova) from a 
$\sim 10 M_\odot$ star in a region containing $\sim 20 M_\odot$ stars.
However, the local stellar environment of SN~2002bu, like that of 
SN~2008S (\citealt{Szczygiel2012}), appears to contain no stars more
massive than $\sim 10M_\odot$.  In fact, the environment of SN~2002bu appears
to favor still lower masses, closer to $ 5M_\odot$ than $10 M_\odot$.  This
strongly favors the interpretation of \cite{Prieto2008a} and \cite{Thompson2009} 
that the SN~2008S class of transients is associated with AGB stars.  The very
low mass scale implied by the environment of SN~2002bu does not, however, 
favor the possibility that the transients are electron capture supernovae
(\citealt{Thompson2009}, \citealt{Botticella2009}),
as these are only expected for masses of $8$-$10 M_\odot$ (e.g. \citealt{Poelarends2008}).
Obtaining deep optical images of the SN~2008S and 2002bu fields to better constrain
the extinction and temperatures of the nearby stars would greatly improve the
characterization of the stellar populations over our present near-IR results.
 
As with all the ``supernova impostors'', the only way to determine their natures is
to continue to track the evolution of their spectral energy distributions.  It
is clear from our survey in \cite{Kochanek2012} that there are at least two 
classes of sources and that no sources with adequate data were consistent with a 
brief period of high mass loss during the transient leading to the formation of
a dusty expanding shell that obscured the source.  The strength of this 
conclusion is strongly limited by the fragmentary nature of the data in
both wavelength and time.  Here we see that the SN~2008S transient
class can remain dust-obscured for long periods of time, primarily because the
optical depth is not evolving like that of an expanding shell.  However, if 
object A is the near-IR counterpart of SN~2002bu, the veil is beginning to thin.

\acknowledgements 

The authors are supported in part by NSF grant AST-0908816.
The authors would like to thank the Swift team and its PI, N. Gehrels,
for approving the Swift TOO observation of SN~2002bu, and J.A. Beacom,
J.-L. Prieto, K.Z. Stanek and T.A. Thompson for their comments.
This work is based in part on observations made with the Spitzer Space
Telescope, which is operated by the Jet Propulsion Laboratory, California 
Institute of Technology under a contract with NASA. Support for this work was 
provided by NASA through award 1414623 issued by JPL/Caltech.
Support for HST program GO-12331 
was provided by NASA through a grant from the Space Telescope
Science Institute, which is operated by the Association
of Universities for Research in Astronomy, Inc., under
NASA contract NAS5-26555.

{\it Facilities:} \facility{HST, SST, Swift}

\vfill\eject

\begin{sidewaystable}
\tabletypesize{\scriptsize}
\begin{center}
\caption{Photometry of SN~2002bu \label{tab:mags}}
\begin{tabular}{cccccccccccccccc}
\tableline
Date (UT) & MJD & PI/Program & $F435W$ & $F555W$ & $F625W$ & $F814W$ & $F110W$ & $F160W$ & $[3.6]$ & $[4.5]$ & $[5.8]$ & $[8.0]$ & $[24.0]$ & $[70.0]$ & Comments\\
          &     &            &  [mag]  &  [mag]  &  [mag]  &  [mag]  &  [mag]  &  [mag]  &[$\mu$Jy]&[$\mu$Jy]&[$\mu$Jy]&[$\mu$Jy]&[$\mu$Jy] & [$\mu$Jy]&          \\
\tableline
2004-04-25 & 53120 & Fazio/69         & \nodata & \nodata & \nodata & \nodata & \nodata & \nodata & \nodata & \nodata  & \nodata   & \nodata   &$750\pm26$& $<4200$ & 1 \\
2004-05-02 & 53128 & Fazio/69         & \nodata & \nodata & \nodata & \nodata & \nodata & \nodata &$544\pm7$& $806\pm9$&$1098\pm12$&$1204\pm20$& \nodata  & \nodata & 1 \\
2005-03-20 & 53456 & Filippenko/10272 &$<25.67$ &$<25.25$ &$<24.83$ &$<24.40$ & \nodata & \nodata & \nodata & \nodata  & \nodata   & \nodata   & \nodata  & \nodata & 1 \\
2011-06-23 & 55735 & Kochanek/80015   & \nodata & \nodata & \nodata & \nodata & \nodata & \nodata & $8\pm3$ & $11\pm1$ & \nodata   & \nodata   & \nodata  & \nodata &   \\
2012-01-31 & 55958 & Kochanek/80015   & \nodata & \nodata & \nodata & \nodata & \nodata & \nodata & $5\pm1$ & $13\pm1$ & \nodata   & \nodata   & \nodata  & \nodata &   \\
2012-02-15 & 55972 & Kochanek/12450   & \nodata & \nodata & \nodata & \nodata&$25.64\pm0.10$& $24.85\pm0.13$&\nodata&\nodata&\nodata& \nodata   & \nodata  & \nodata & A \\
           &       &                  &         &         &         &       &$27.21\pm0.39$& $26.72\pm0.71$&       &       &       &           &          &         & B \\
           &       &                  &         &         &         &       &$25.78\pm0.12$& $25.53\pm0.26$&       &       &       &           &          &         & C \\
\tableline
\end{tabular}
\tablecomments{All the magnitude upper limits are $3\sigma$.
 The date of discovery of the transient is MJD $52392.3$.
 (1) From \cite{Kochanek2012}; (A) measurement for Source A; (B) measurement for Source B; (C) measurement for Source C.}
\end{center}
\end{sidewaystable}

\vfill\eject

\begin{deluxetable}{lccccccl}
\tabletypesize{\scriptsize}
\tablecaption{Spectral Energy Distribution Models \label{tab:results}}
\tablehead{
MJD  & $\log_{10} T_*$ & $\log_{10} L_*$    & $\log_{10} T_d$  &$\log \tau_V$ &$\log_{10} R_{in}$ &$\log_{10} v_s$   &Comment\\
  & (K)   & ($L_\odot$)  & (K)    &      & (cm)    &(km/s)  \\
 }
\startdata
53128 &$3.99\pm0.31$ &$4.92\pm0.01$ &$3.04\pm0.07$ &$1.58\pm0.21$ &$15.77\pm0.13$ &$2.94\pm0.13$ &Graphitic \\
53128 &$3.88\pm0.32$ &$4.95\pm0.02$ &$3.09\pm0.10$ &$1.77\pm0.27$ &$15.62\pm0.07$ &$2.80\pm0.07$ &Silicate \\
55972 &$4.32\pm0.09$ &$4.52\pm0.22$ &$2.65\pm0.05$ &$0.70\pm0.25$ &$16.18\pm0.25$ &$2.72\pm0.25$ &Graphitic  \\
55972 &$4.43\pm0.04$ &$4.95\pm0.18$ &$2.77\pm0.04$ &$1.21\pm0.11$ &$15.98\pm0.17$ &$2.52\pm0.17$ &Silicate  \\
\enddata
\tablecomments{These include a weak prior on the stellar temperature, $\log_{10} T_* = 4.0 \pm 0.3$ ($3000<T_*<30000$), 
  and the velocity $\log_{10} (v_s/\hbox{km/s}) =\log_{10} 893 \pm 0.3$ at which $R_{in}$ expands.  
   These models all assume that
  source A is the near-IR counterpart of SN~2002bu.  If it is a chance coincidence, then the optical depth scale
  can be significantly higher.  
  }
\end{deluxetable}

\end{document}